# Techno-economic analysis of Power-to-Gas plants in a gas and electricity distribution network system with high renewable energy penetration

Gabriele Fambri[a], Cesar Diaz-Londono[a], Andrea Mazza[a], Marco Badami[a], Teemu Sihvonen[b], Robert Weiss[b]

[a] *Dipartimento Energia, Politecnico di Torino, Corso Duca degli Abruzzi 24, 10129 Torino, Italy*
[b] *VTT Technical Research Centre of Finland Ltd., P.O. Box 1000, 02044 VTT, Espoo, Finland*

## HIGHLIGHTS

- The utilization of P2G plants in electricity and gas distribution networks has been analyzed.
- Medium pressure gas network can be used as an intraday storage means for SNG.
- A low gas demand implies coordinated P2G operation to optimize the whole multi energy system.
- Only in the best-case scenario is the SNG cost comparable with the cost of natural gas.

## ABSTRACT

*Distributed generation, based on the exploitation of Renewable Energy Sources (RES), has increased in the last few decades to limit anthropogenic carbon dioxide emissions, and this trend will increase in the future. However, RES generation is not dispatchable, and an increasing share of RES may lead to inefficiencies and even problems for the electricity network. Flexible resources are needed to handle RES generation in order to support the delicate electricity generation and demand balance. Energy conversion technologies (P2X, Power to X) allow the flexibility of energy systems to be increased. These technologies make a connection between different energy sectors (e.g., electricity and gas) possible, and thus create new synergies within an overall multi-energy system. This paper analyzes how the P2G technology can be used at the distribution network level (both gas and electricity) to optimize the use of RES. In fact, in order to coordinate P2X resources, it is necessary to take into account the whole multi-energy scenario, and not just the electrical side: it therefore becomes fundamental to recognize the pros and cons that Balancing Service Providers (BSPs), composed of a number of P2G plants (representing the Balancing Responsible Providers, BRPs), may have when offering services to an electricity network. Moreover, the convenience of the decarbonization of the gas grid has been evaluated through the calculation of the levelized cost of Synthetic Natural Gas ($LC_{SNG}$) for cost scenarios for the years 2030 and 2050, considering different assumptions about the cost of the surplus utilization of RES. The results show that $LC_{SNG}$ may vary from 47 to 319 €/MWh, according to the different configurations, i.e., only in the best-case scenario is the SNG cost comparable with the cost of natural gas, and hence does the P2G technology result to be profitable.*

## ACRONYMS

| | |
|---|---|
| AC | Alternating Current |
| BRP | Balance Responsible Party |
| BSP | Balancing Service Provider |
| CAPEX | Capital Expenditure |
| $CO_2$ | Carbon dioxide |
| DC | Direct Current |
| DSO | Distribution System Operator |
| EN | Electricity Network |
| G2P | Gas-to-Power |
| GN | Gas Network |
| $H_2$ | Hydrogen |
| HV | High Voltage |
| $LC_{SNG}$ | Levelized Cost of SNG |
| LHV | Lower Heating Value |
| MV | Medium Voltage |
| NG | Natural Gas |
| $O_2$ | Oxygen |
| OPEX | Operational Expenditure |
| P2G | Power-to-Gas |
| $P2H_2$ | Power-to-Hydrogen |
| P2X | Power-to-X |
| PEM | Polymer Electrolyte Membrane electrolyzer |
| RES | Renewable Energy Sources |
| RPF | Reverse Power Flow |
| STP | Standard Temperature and Pressure |
| PV | Photovoltaic plants |
| SNG | Synthetic Natural Gas |
| TR | Transformer |
| TSO | Transmission System Operator |
| WT | Wind Turbines |
| WACC | Weighted Average Cost of Capital |

## 1. Introduction

Almost the entire scientific community agrees that anthropogenic carbon dioxide ($CO_2$) emissions have led to an increase in the world's mean temperature of 0.8 °C since the end of the nineteenth century [1]. In order to mitigate this dangerous trend, the European Union, with the "Clean Energy for all Europeans" legislation package [2], has defined rules to achieve carbon neutrality by 2050. One of the main goals of this package is an increase in the renewable share, in order to dismiss the utilization of fossil fuels. As a result of the intrinsic non-dispatchability, high volatility and intermittency nature of renewable generation, the increase in the share of renewable energy should be supported by an increase in the flexibility of the system, in order to balance electrical consumption and generation [3],[4]. The installation of such electric storage technologies as electric batteries [5], pumped hydro storage [6] and compressed-air energy storage [7] can partially provide the required flexibility: these technologies make it possible to absorb excess renewable energy production, and to release the accumulated energy when necessary. Electricity storage is characterized by a very high investment cost; moreover, as concluded in [8], the optimal solution cannot be found by considering only a single part of the overall energy system. To tackle this problem more efficiently, the paradigms of the management of the overall energy system need to be reviewed: a more holistic approach, which includes the interactions of the electricity sector with other energy sectors, allows new and non-negligible sources of flexibility to be exploited [9],[10]. In fact, other non-electrical energy sectors are usually more flexible than the power sector, because they do not require an instantaneous balance between load and generation. Hence, thanks to certain energy conversion technologies (Power-to-X, P2X), it could be possible to exploit the synergy of multi-energy systems and use their inherent flexibility to offer ancillary services to the electricity grid [11], by allowing an increasing share of Renewable Energy Sources (RES) to be accommodated within the electrical system. One of the most frequently discussed P2X technologies is the so-called Power-to-Gas (P2G) technology. The term P2G can be used to indicate both Power-to-Hydrogen ($P2H_2$) technologies (where the electricity is used to produce hydrogen as the final product) and Power-to-Methane technologies (where the production chain also includes a methanation unit to transform hydrogen into Synthetic Natural Gas, SNG). Even though the Power-to-Methane technology needs an additional conversion step to the Power-to-Hydrogen technology, with a consequent reduction in the overall efficiency, it offers a number of advantages: a) the SNG volume energy density (> 1000 kWh/m$^3$) is much higher than that of hydrogen (270 kWh/m$^3$) [12]; b) SNG can be injected into the existing gas infrastructure, while hydrogen can only be injected at low concentrations, due to hydrogen embrittlement, which can create cracks in iron and steel pipes [13]; c) hydrogen has a higher risk of ignition than SNG, thus making it less safe for domestic utilization [13]; d) producing SNG is also advantageous to promote $CO_2$ capture technologies and favor its utilization value chain. Accordingly, hydrogen can be mixed with carbon dioxide obtained from several sources (e.g., flue gases, biogas, air) and stored in synthetic hydrocarbons [14]. This paper focuses only on Power-to-Methane: hence, hereinafter P2G will be used to indicate the Power-to-Methane technology.

The utilization of P2G plants connected to transmission networks has been widely studied. In [15], the optimum P2G size was defined for a regional scenario. In [16] it was shown how an industrial P2G plant connected to the transmission grid could optimally operate simultaneously in both wholesales energy and ancillary service markets, and strong impacts of the national differences in market rules were pointed out. A dimension optimization in [17] showed how a P2G plant could operate on energy and ancillary service markets while absorbing contracted local wind and solar power to avoid reverse power flow. Oversizing the electrolyzer compared to methanation was significantly beneficial, increasing the share of the local renewable power in the product SNG. In [18], it was shown that the coordination of flexible resources, including P2G, reduces wind curtailment and increases social welfare. In [19] it was shown how to reach cost-efficiently a 100% decarbonized urban power- and district heating system for a large city using wind and solar power, requiring either city level P2G solutions or multiple reinforcement of the transmission power grid. The role of the P2G technology in a near zero carbon footprint European scenario for the year 2050 was analyzed in [12]. In [20], a gas transmission system was used to store the surplus generation of RES.

However, as was also concluded in [21]–[23], only a few studies have focused on the integration of P2G with the distribution system. The presence of just a few studies about the integration of P2G with the distribution system may partially be explained by considering the low level of maturity of the technology, which makes it difficult to suggest models of the entire P2G chain that are suitable for integration with network calculations. In [24], the authors investigated the possibility of absorbing the excess energy of PV in the distribution network for a region in Southern Germany using hundreds of small-scale P2G installations (300-700 kW electrolyzers). The voltage control of a power distribution network was analyzed in [25] using an On-Load Tap Changer and an alkaline $P2H_2$. The same $P2H_2$ was then used in [26] to analyze the optimum size and allocation of a plant in order to reduce the impact of an increasing RES installation on the distribution network. In [22], the installation of an electrolyzer in an electricity distribution network was evaluated for the absorption of the excess production of PV, in comparison to a network expansion solution. In [27] the interaction between the gas network and the local distribution networks for electricity and district heat were analyzed for a small town to show how to reach a 100% wind and solar power based urban electricity and heating system in cold climate regions, enabled by the gas grid and P2G. In [28], a technoeconomic analysis of a P2G plant was carried out considering different configurations, the optimal P2G capacity was defined, and P2G was also analyzed considering the integration of electrochemical storage to improve the continuous operation of the plant. In [23],[29], the authors focused on the utilization of P2G and Gas-to-Power (G2P) technologies for voltage regulation in the distribution network and presented a new algorithm for real time scheduling. In [21], the authors analyzed the benefits of a P2G plant in a distribution network with high-RES penetration and presented a novel model of a P2G plant based on real data.

The role of distributed resources, connected to the distribution system, is gaining increasing importance for the control and regulation of power systems: in fact, their increasing penetration calls for their participation on both the energy and ancillary service markets, in the latter case to support the proper operation of electrical systems. In this context, the European Union has recently opened new perspectives for all distributed resources, by fostering their participation, as aggregated sources, on the ancillary service market [30]. The market players, called Balancing Service Providers (BSP), are able to offer either a balanced energy or balanced capacity to the market: each BSP may aggregate a number of generators or demand facilities, and they can thus offer *flexibility* to Transmission System Operator (TSO). It is worth noting that, in the future, this form of flexibility may be offered to Distribution System Operators (DSO), to solve local issues that affect their networks, as indicated, for example, in the SmartNet project [31]. Once a BSP gets a bid accepted, it is assigned to one or more Balance Responsible Parties (BRPs), who are the final parties responsible for maintaining a balance at the connection node. The BRP is the owner of the flexible facility, or its representative. This paper, which is based on these premises, considers three P2G plants (i.e., three BRPs), which are managed together (i.e., through a strategy somehow defined by a BSP) to solve the local problems that affect the distribution system, such as the reverse power flow (RPF). However, as a result of the nature of the P2G plants, and unlike the previously cited literature, the paper conducts an analysis that takes into account both the electricity network and the constraints imposed by the dynamics of P2G plants and the gas network. This may open new perspectives, because it may lead to the concept of *multi-energy prosumer* (or *global prosumer*) [32], in which the gas grid becomes *active*, and the single prosumer may act as either a producer or a consumer in both electrical and gas systems. In such a scenario, it is necessary to verify that the injection of SNG into the gas system does not affect the system constraints, i.e., this means that the prosumer also plays the role of BRP for the gas system. In the specific case dealt with in this paper, there is no global prosumer, because there are no distributed generation plants associated with the P2G facilities; however, the consideration about the role of BRP for both electrical and gas grids is still valid. The interaction between electricity network, gas network and the P2G components has been analyzed in this paper. In order to create a critical case study, a scenario with a high RES generation and a high seasonal influenced gas demand has been considered. The scenario is characterized by periods of high RES over-generation and low gas demand. In such a scenario, the different P2G plants operate as separate BRPs under the same BSP. It has been analyzed how, in order to reach the optimal condition of the overall multi-energy system, the different BRPs need to be coordinated not only just considering the power sector alone. The scenario has been analyzed from an economic point of view by calculating the Levelized Cost of SNG ($LC_{SNG}$) for each P2G plant.

The remainder of this paper is structured as follows. Section 2 introduces the analyzed case study, the simulation input data and the models used in this study; the simulation results are described in Section 3, whereas Section 4 summarizes the conclusions.

## 2. Models and methodological frameworks

The analyzed multi-energy system includes both electricity and gas distribution networks, coupled thanks to the utilization of P2G technologies (see Fig. 1). The district gas and electricity scenario has been developed and simulated for one year with a time resolution of 15 minutes. In the developed scenario, the P2G plants act as BRPs, part of a portfolio under the same BSP coordination. The P2G plants are used to mitigate the energy unbalances caused by RES connected to the distribution electricity network. In fact, thanks to an energy conversion process, P2G plants make it possible to exploit the flexibility and storage capacity of a gas network and transpose it to the electricity sector. As shown in Fig. 1, the operation of P2G BRP plants is a function of the electricity network conditions, and in particular of the presence of RES over-generation, but also of the availability of the gas network to receive SNG without violating the operating pressure limits, or the state of the plant itself. More details about the coordination logic are reported in Section 2.6.

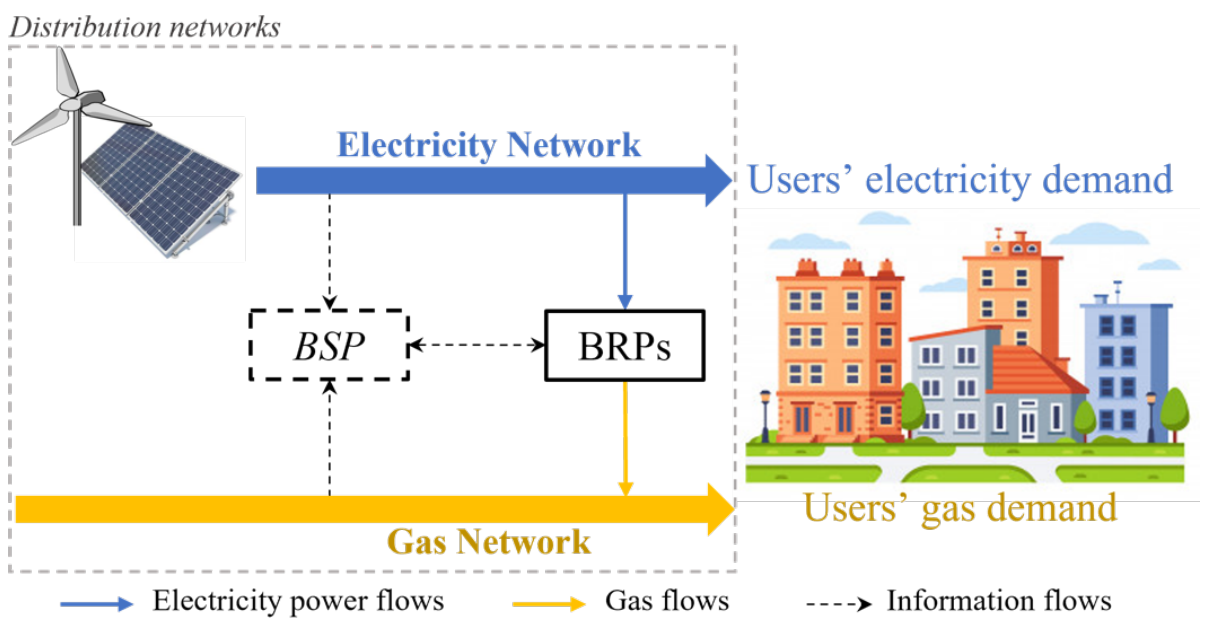

Fig. 1. Flow chart of the considered multi-energy system model.

## 2.1. Electrical system

A sample of the urban distribution system of a city in northern Italy has been used for the electricity network (EN). The network is composed of 43 nodes supplied by three HV/MV transformers (see Fig. 2). As depicted in Fig. 2, different RES plants, in particular photovoltaic (PV) and wind turbines (WTs), are spread over the network.
The network is solved through the Backward Forward Sweep method [33], development in Simulink, as shown in [34].

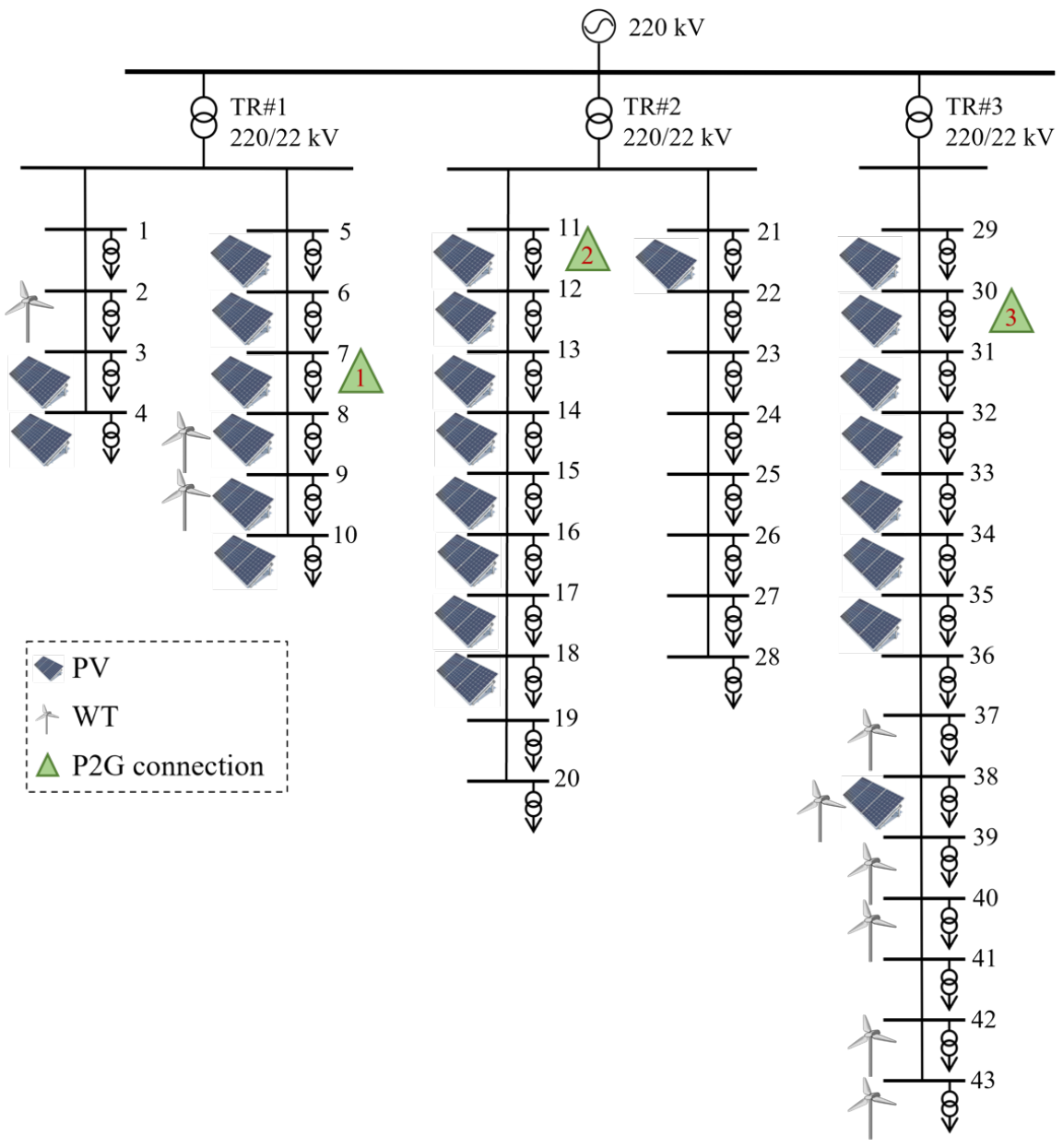

Fig. 2. Topology of the electricity network.

The scenario considers the presence of RPF, which may affect both the transmission and the distribution system. If an RPF exists in only one or two feeders, it is totally or partially redirected to the other ones through the busbar that connects all the feeders. In this case, the transmission system is not affected (or is only affected very slightly) by the consequences of the RPF, and hence the RPF introduces local issues (for example, the coordination of protection systems) into the distribution system. Conversely, if an RPF exists in all the feeders, it is necessarily injected into the transmission system, thus causing dispatching issues at the TSO as well. In this analysis, BSP is used to offer ancillary services to the DSO and thus to balance the RPFs of each HV/MV transformer in the distribution system.

## 2.2. Gas system

The topology of the gas network (GN) (see Fig. 3) was derived from the one presented in [35]: a medium-pressure distribution network (4th species according to the Italian DM 24/ 11/1984 classification [36]) that covers an urban area of around 29 km$^2$. The pressure of a 4th species gas network needs to stay within a range of 1.5 – 5 bar$_g$. The medium-pressure network is connected to a high-pressure network through a citygate (node 1), and at this point the gas is injected from the high-pressure network into the medium-pressure network (the gas cannot flow from the medium-pressure network to the high-pressure network). The gas is supposed to be injected at 4 bar$_g$. The gas network is divided into nodes: each withdrawal, injection and junction point is considered as a node (see Fig. 3). The model calculates the pressure evolution in each node and the gas flow in each pipe of the network. More information about the gas network model and its validation are reported in the Appendix.

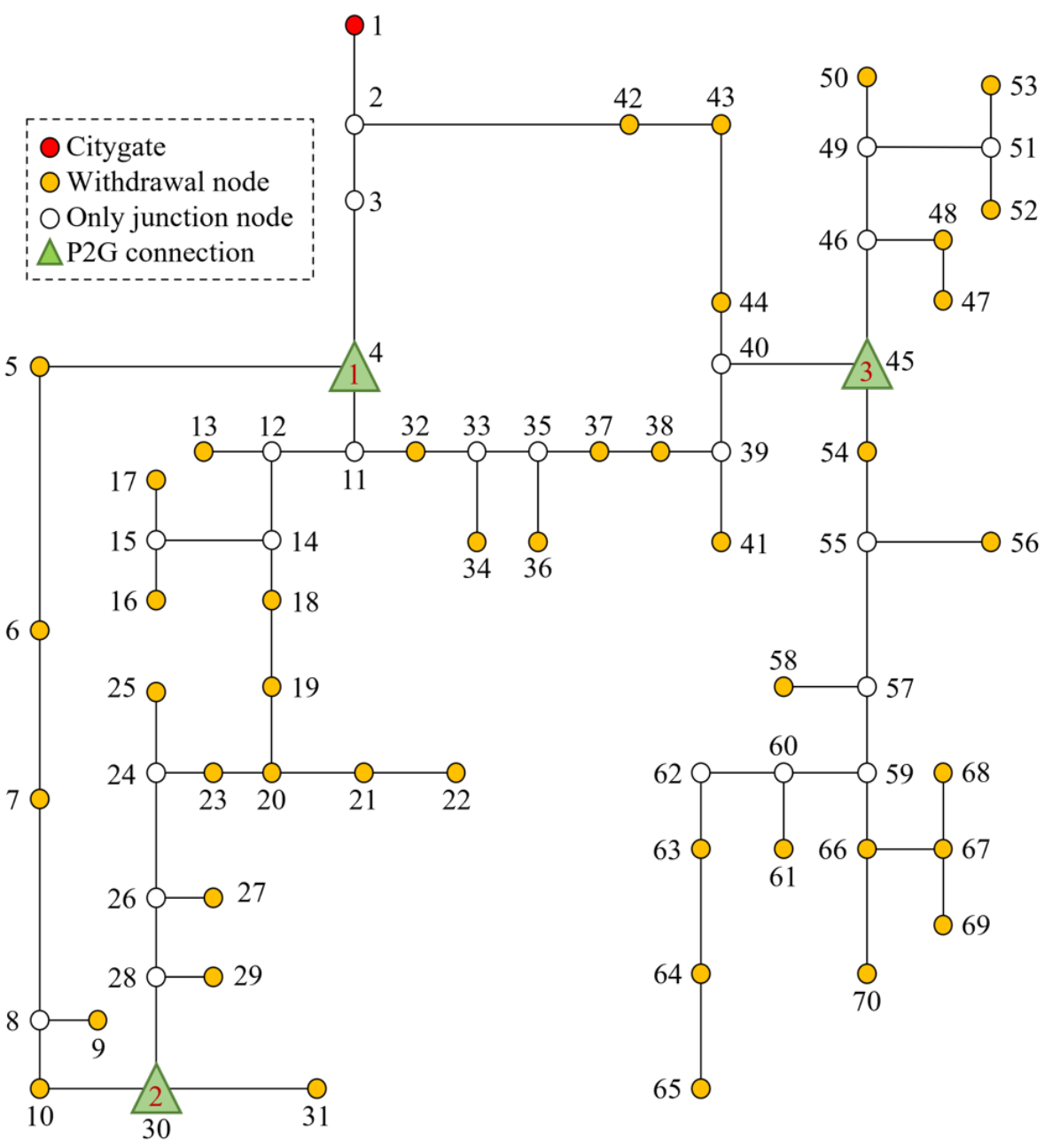

Fig. 3. Topology of the gas network.

## 2.3. Power-to-Gas systems

The three P2G plants have the same size and characteristics. Each plant has a Polymer Electrolyte Membrane (PEM) electrolyzer of 1200 kW (in terms of electricity input), and a catalytic methanation reactor of 600 kW (in terms of SNG output). As schematized in Fig. 4, the PEM electrolyser consumes electricity to produce hydrogen at a pressure of 30 bar (commercial state-of-the art), which is directly fed into an internal hydrogen buffer to be accumulated and used by the methanation unit to produce SNG. In the model configuration, the hydrogen buffer was selected to host up to 92 kg of hydrogen, corresponding to 1024 m$^3$ STP (Standard Temperature and Pressure) and in terms of energy, to 3060 kWh considering the Lower Heating Value (LHV) of hydrogen. The PEM electrolyser model is based on an lumped linear operational efficiency curve, which is the hydrogen output in relation to the electrical power input, combining power-to-hydrogen conversion efficiency in the stack, the transformer and rectifier efficiencies, as well as the balance-of-plant power consumption, while part of the conversion losses are modeled as heat generation that can be recovered [27]. The catalytic methanation unit model is a surrogate model that has been derived from the simulation data provided from a high-fidelity [37],[38] based on Apros® dynamic process simulator [39]. Earlier versions of the electrolyser and methanation surrogate models have been utilized in transmission network level studies on P2G plant integration to different power, regulation and ancillary service markets [16],[17], and further details on the methanation surrogate model can be found in [40].
Since the simulation time step is 15 minutes and the PEM can vary its operating setpoint in seconds, the ramp-up and ramp-down constraints have been neglected. The methanation unit has a lower dynamic: according to results from the high-fidelity P2G process model of a 600kW methanation reactor configuration, the 600 kW methanation unit can increase its hydrogen consumption by about 3.8 kg per hour and decrease it by 46 kg per hour while sustaining the specified SNG product gas quality without violating a maximum hydrogen content limit (set here to <4%) required e.g., by the gas network operator. The methanation unit cannot work below 50% of its nominal capacity, without requiring shutdown. The P2G technical parameters are summarized in Table 1.

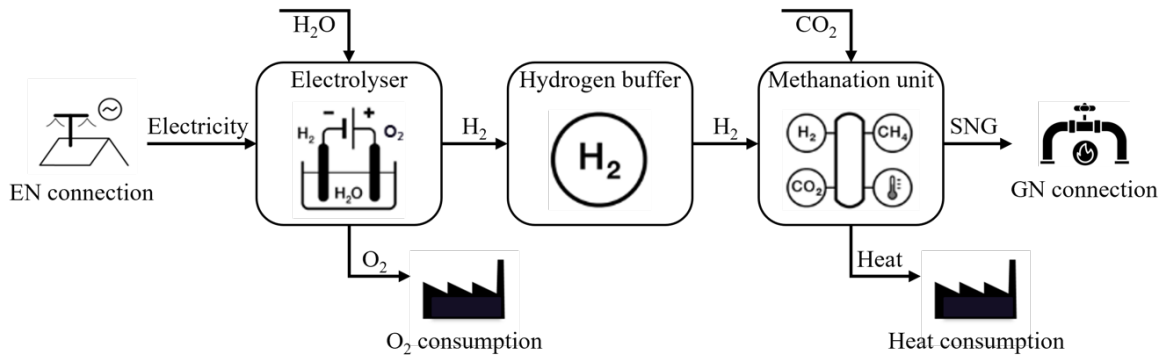

Fig. 4. Scheme of P2G system.

Table 1. Technical parameters of the P2G plants.

| Parameter | Unit | Value |
|---|---|---|
| Electrolyzer capacity | kW (el. input) | 1200 |
| Meth. unit capacity | kW (SNG output) | 600 |
| Hydrogen buffer capacity | m$^3$ of $H_2$ (at STP) | 1024 |
| Meth. unit - minimum load | % | 50 |
| Meth. Unit - max ram up | kg ($H_2$ input) / h | 3.8 |
| Meth. unit - max ram down | kg ($H_2$ input) / h | 46 |

The position of the P2G plants was defined according to the methodology shown in [41]. Each P2G is connected downstream of a different HV / MV transformer. This correspondence allows the RPFs on each transformer to be absorbed by one of the three P2G plants. The connection to the gas network is made in order to distribute the various connections over the network. The P2G connections to the distribution networks are summarized in the Table 2.

Table 2. Network connections to the P2G plants.

| | EN node connection | GN node connection |
|---|---|---|
| P2G#1 | 7 (TR#1) | 4 |
| P2G#2 | 11 (TR#2) | 30 |
| P2G#3 | 30 (TR#3) | 45 |

### 2.4. Operation scenario

A critical energy context was chosen to investigate P2G operation. The scenario is characterized by a high number of RES penetrations, which are not equally distributed over the electricity network, i.e., the RES plants are more concentrated downstream of the third HV/MV transformer (see Table 3). Moreover, the scenario is assumed to mainly feed residential and tertiary sector users. Thus, the natural gas demand is affected seasonally to a great extent as a result of the high gas demand for building heating purposes during winter: the gas consumption during the heating season (from 1st Jan. to 15th Apr. and from 15th Oct. to 31st Dec.) is about 10 times higher than during the rest of the year. The electricity user demand capacity, the total RES installation (for each transformer and for the whole network) and the peak gas demand are summarized in Table 3. The duration curves of the electricity demand, gas demands and renewable production are shown in Fig. 5a, while Fig. 5b shows the renewable production and energy demands per month. The demand for electricity is roughly constant throughout the year. RES production increases considerably in the summer months due to the influence of solar radiation (the renewable production in the summer months is almost double that of the winter months). Thus, the winter season is characterized by a contained excess of RES and a high gas demand, while in the summer, when the highest RES overproductions occur, the natural gas demand is much lower.

Table 3. Electricity demand, gas demand and RES installations.

| | Installed power [MW] | | | |
|---|---|---|---|---|
| | TR#1 | TR#2 | TR#3 | Total |
| EL demand | 3.90 | 9.30 | 5.10 | 12.30 |
| PV | 3.90 | 4.50 | 6.60 | 14.30 |
| WT | 0.80 | 0 | 3.60 | 4.40 |
| | Peak demand [MW] | | | |
| Gas demand | 23.0 | | | |

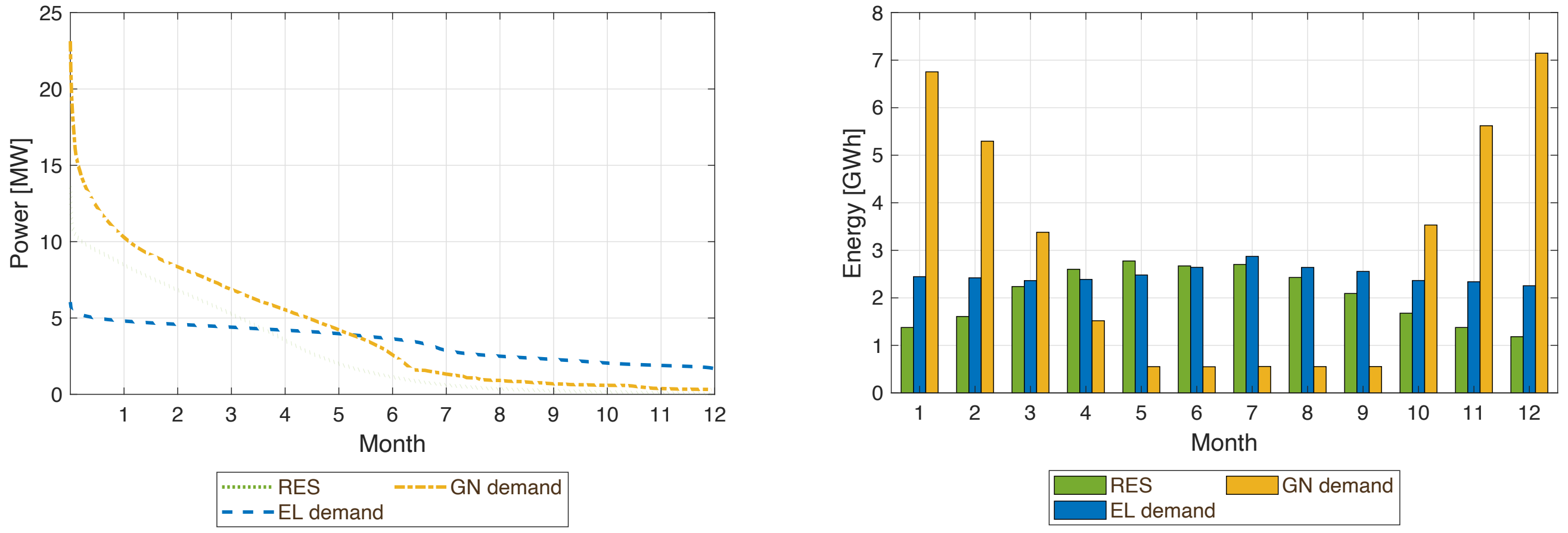

(a) (b)

Fig. 5. RES generation, electricity demand and natural gas demand duration curves (a). Monthly RES generation, monthly electricity and gas demands (b).

### 2.5. Economic assumptions

#### 2.5.1. Capital and fixed operational expenditure

The cost of P2G shows a decreasing trend that will most probably continue, especially if the P2G plant components are manufactured in standardized sizes and series [42]. Two economic scenarios have been considered for this study: the first one refers to the year 2030, whereas the second one pertains to the year 2050. The assumed capital expenditure (CAPEX) and the fixed operational expenditure (OPEX) are shown in Table 4.

Table 4. CAPEX, OPEX and lifetime of the main P2G components.

| | Unit | CAPEX [€/unit] | | OPEX [% of CAPEX] | | Ref |
|---|---|---|---|---|---|---|
| | | 2030 | 2050 | 2030 | 2050 | |
| Electrolyzer | $kW_e$ | 650 | 400 | 3 | 2 | [13],[42]–[44] |
| $H_2$ buffer | $m^3$ $H_2$ | 75 | 50 | 1.5 | 1.5 | [42],[45] |
| Methanation unit | $kW_{SNG}$ | 500 | 300 | 5 | 3 | [42],[46],[47] |

#### 2.5.2. Variable operational expenditures and revenues

The P2G plant consumes electricity to feed the electrolyzer which consists of the direct current (DC) power consumed by the water splitting to hydrogen and oxygen in the electrolyser stacks, the conversion losses in the transformer and rectifier to supply the needed DC power, as well as auxiliary alternating current (AC) AC power needs for the balance of plant (such as electrolyser and methanation unit feed pumps, hydrogen dryers, automation and control system, etc). A commercial state-of-the-art PEM electrolyser is capable of supplying output hydrogen at a 30 bar pressure, which is assumed to be the maximum pressure level of the intermediate hydrogen storage, and thus no hydrogen compressor is assumed to be needed. The scenario is characterized by electricity surplus periods (where RES over-generation occurs) and electricity deficit periods. Hence, the price of the electricity has been considered different for the two cases [28]. The cost of the deficit electricity has been considered equal to 60 €/MWh [28], while an incentive cost has been considered for the surplus period, after evaluating different prices (from 0 to 30 €/MWh). This is due to the fact that the utilization of surplus energy results in a positive externality, because it allows problems related to RES over-generation to be mitigated.

The stack needs to be replaced for the maintenance of the PEM electrolyzer. The replacement cost is assumed to be 35% of the total PEM capital cost, and the replacement takes place with a frequency of once every 5 years [48]. The cost for the $CO_2$ used for the methanation process has been considered equal to 70 €/t [49]. The cost for the demineralized water has been neglected, as it has a marginal impact on the overall costs [42]. Oxygen and high temperatures are generated during P2G operation as indirect products. It is assumed that these products are valorized with a profit of € 80 per ton of oxygen and € 30 per MWh of generated heat. The economic data are summarized in Table 5.

Table 5. Economic assumptions for the considered P2G.

| Parameter | Unit | Value | Ref |
|---|---|---|---|
| Plant lifetime | y | 20 | [42] |
| Deficit electric energy price | | 60 | [28] |
| Surplus electric energy price | | 0-30 | Our assumption |
| PEM replacement | % of CAPEX | 35 | [48] |
| $CO_2$ specific cost | €/t | 50 | [49] |
| $O_2$ specific revenue | €/t | 70 | [49],[50] |
| Heat specific revenue | €/MWh | 30 | [49] |

#### 2.5.3. Levelized cost of SNG

The P2G systems are evaluated from an economic point of view by calculating the levelized cost of SNG ($LC_{SNG}$). $LC_{SNG}$ is the breakeven selling price of the produced SNG, calculated as [51]:

$$LC_{SNG} = \frac{\sum_{i=0}^{n} \frac{CAPEX_i + OPEX_i + C_i^{el} + C_i^{CO_2} + C_i^{R} - R_i^{O_2} - R_i^{heat}}{(1+WACC)^i}}{\sum_{i=0}^{n} \frac{E_i^{SNG}}{(1+WACC)^i}} \quad (1)$$

where:

- $n$ is the lifetime of the plant;
- $CAPEX_i$ is the capital expenditure for all the components (considered only at year 0);
- $PEX_i$ is the fixed operational expenditure for all the components in year *i*;
- $C_i^{el}$ is the cost of the purchased electricity in year *i*;
- $C_i^{CO_2}$ is the cost of $CO_2$ purchasing in year *i;*
- $C_i^R$ is the cost of the replacement of the stack of the electrolyzer (considered only for years 5, 10 and 15);
- $R_i^{O_2}$ is the revenue for the oxygen produced in year *i;*
- $R_i^{heat}$ is the revenue for the high temperature heat produced in year *i;*
- $WACC$ is the Weighted Average Cost of Capital, assumed to be equal to 8% [51];
- $E_i^{SNG}$ is the total amount of SNG energy produced in year *i.*

2.6. Simulation control algorithm

The three P2G plants are connected downstream of three different transformers (see Fig. 2): for this reason, the operation of each P2G only affects the portion of the network it is connected to. For example, P2G 1 cannot be used to absorb the RPF from transformers 2 and 3 but only that from transformer 1.
If there is no energy overproduction downstream of the transformer, the electricity consumption of the P2G system is kept at its minimum load: that is, for the electricity consumption of the auxiliary components and to keep the system in hot standby. The electrolyzer is turned on when an RPF occurs. The upper limit setpoint is determined by considering the nominal power of the plant and the availability of the buffer. If the hydrogen buffer has reached its maximum operating pressure, the electrolyzer cannot operate unless the methanation unit is in operation as well; in this case, it can produce an equal amount of $H_2$ to the $H_2$ consumed by the methanation unit, thereby maintaining the pressure of the buffer within its bounds (see Fig. 6).

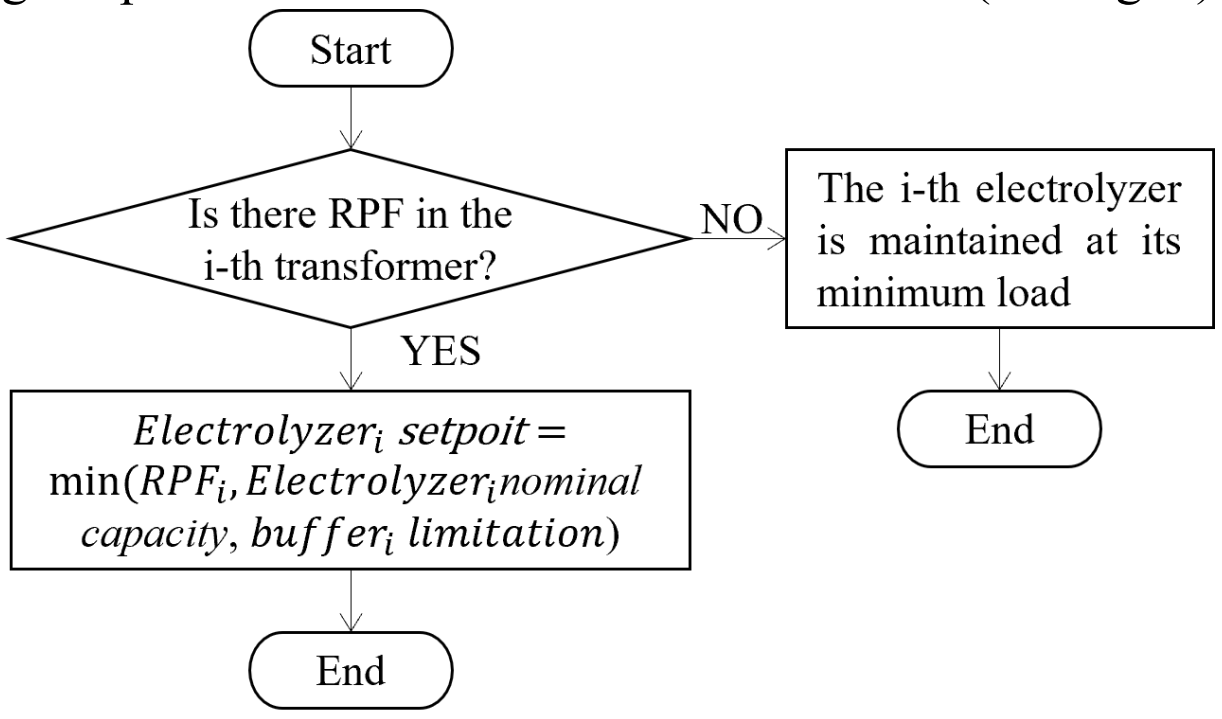

Fig. 6. Control algorithm of the i-th electrolyzer.

If no external restriction occurs, the methanation units are turned on when the hydrogen buffer reaches a pressure of 15 bar. The conversion of $H_2$ into the SNG process continues until the buffer is emptied. However, during low NG demand periods, the GN injection availability could be a constraint for the operation of the methanation unit.
If the SNG injection is limited, the use of plants that are in the up and running mode is prioritized (even though the plants that are not in operation have accumulated more hydrogen). If the GN does not allow operation of all the up and running units, the units with lower amounts of stored hydrogen are set to the stand-by condition. The units in the stand-by condition are only turned on if the up and running units have already reached their maximum load (a limit that is defined by the maximum capacity of the plant and its ramp constraint) and only if the GN allows more SNG to be injected. The methanation unit control logic is summarized in the block diagram in Fig. 7.
It should be noted that, in the case of SNG injection limitations, the various P2G plants should be coordinated to optimize the multi-energy system as a whole. In fact, even though the different BRP systems operate on different portions of the electricity network (such as in the scenario where each plant is downstream of a different transformer), coordination that takes into account the constraint of the common resource is necessary when the systems involve the use of a common flexibility resource (in this case the gas network). In order to maximize the overall benefits, it is necessary to favor the resources that can lead to the greatest benefits for the overall system. For example, the use of the units already in operation is prioritized in low GN flexibility conditions in order to limit the overall number of shutdowns. In the same way, if several systems are in the same state (i.e., all off or all on), the use of the system that has stored the greatest amount of hydrogen inside the buffer is favored in order to increase the continuous operation of the systems as much as possible and, at the same time, increase the availability of use of electrolyzers.

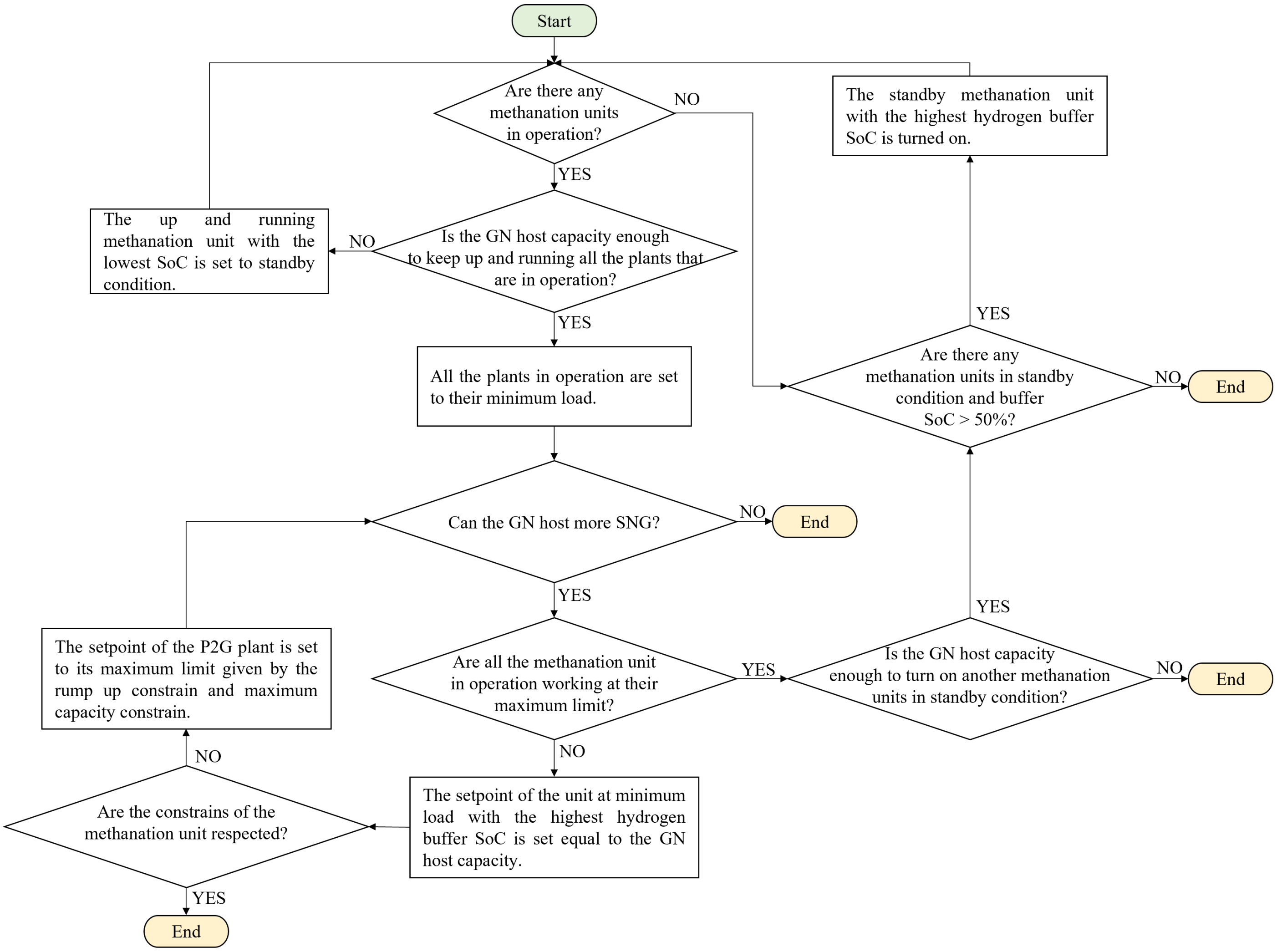

Fig. 7. Control algorithm of the methanation units.

## 3. Results and discussion

### 3.1. Energy impact of P2G on the electricity and gas networks in the winter and summer seasons

Fig. 8 reports the energy flow for the HV/MV transformers of the electricity network for a typical winter day. The renewable energy production causes over-generation in all the HV/MV transformers. Transformer #3 is the one that is affected by the highest RES overproduction (see Table 6). Thanks to the fast response of the PEM electrolyzer, the P2G loads follow the RES surplus and absorb about 85% during the whole heating season. When the RES surplus is higher than the PEM capacity (1.2 MW), the electricity excess can no longer be absorbed, and this causes an RPF of the transformer (see the yellow area in Fig. 8). During the heating season, the gas network can absorb all the produced SNG, without violating any operational constraints for the P2G plants. The SNG production is consumed directly, due to the high gas demand (see Fig. 9a): in this season, the SNG injections covers only 4% of the total gas demand (see Table 7). Fig. 9b reports the evolution of the gas network pressure: it can be noted that the SNG injection does not cause any relevant pressure variation. The figure reports the highest and lowest network pressures for each timestep. The highest pressure is found in the most upstream nodes, where the pressure is kept close to the injection pressure of the citygate (4 $bar_g$). The lowest pressure is recorded in the nodes far from the citygate. In terms of time, the steps with the lowest network pressure correspond to the periods of the highest gas demand.

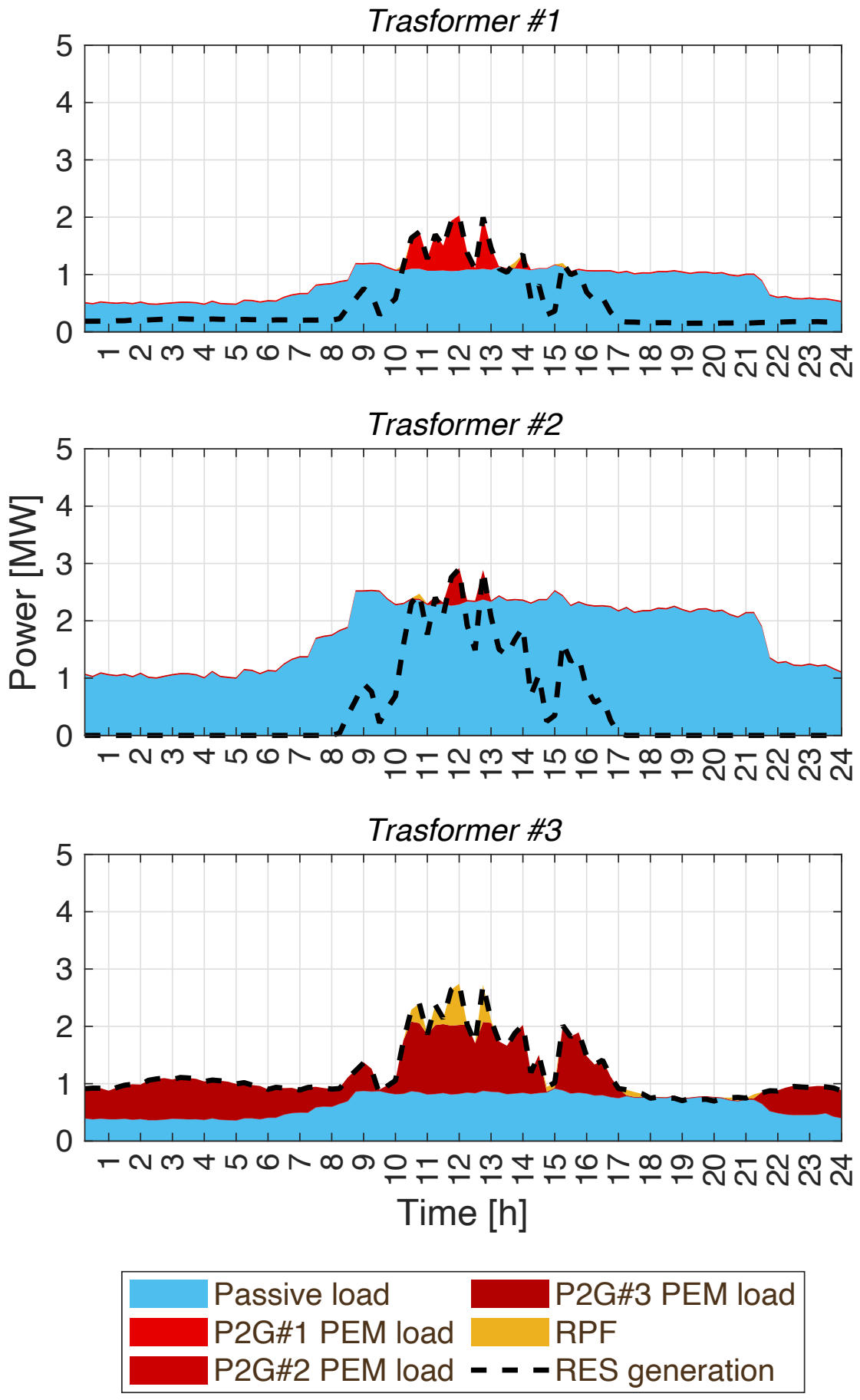

Fig. 8. Balance of the electricity network transformers for the winter.

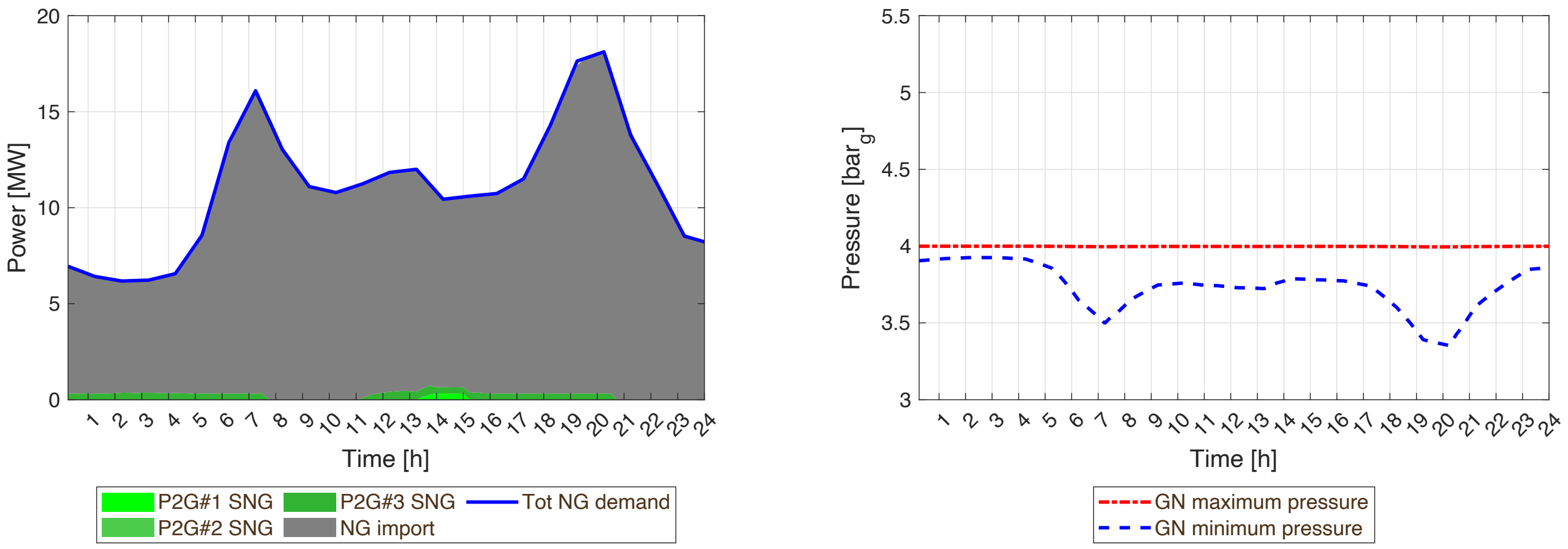

(a) (b)

Fig. 9. Balance of the gas network (a) and gas network pressure (b) for the winter.

Table 6. Electricity network results.

| | Unit | Heating season | | | | No heating season | | | | Whole year | | | |
|---|---|---|---|---|---|---|---|---|---|---|---|---|---|
| | | TR#1 | TR#2 | TR#3 | Tot | TR#1 | TR#2 | TR#3 | Tot | TR#1 | TR#2 | TR#3 | Tot |
| EL demand | GWh | 3.70 | 7.80 | 2.79 | 14.29 | 4.12 | 8.71 | 2.97 | 15.80 | 7.82 | 16.50 | 5.76 | 30.08 |
| RES | GWh | 2.41 | 3.23 | 4.12 | 9.76 | 3.90 | 5.68 | 5.39 | 14.97 | 6.32 | 8.91 | 9.51 | 24.73 |
| Surplus | GWh | 0.67 | 0.46 | 2.04 | 3.17 | 1.40 | 1.08 | 3.08 | 5.55 | 2.07 | 1.54 | 5.11 | 8.72 |
| Absorbed surplus | GWh | 0.62 | 0.40 | 1.61 | 2.63 | 1.23 | 0.90 | 2.07 | 4.20 | 1.86 | 1.30 | 3.68 | 6.83 |
| RPF | GWh | 0.05 | 0.06 | 0.43 | 0.53 | 0.16 | 0.18 | 1.01 | 1.35 | 0.21 | 0.24 | 1.44 | 1.88 |

Table 7. Gas network results.

| | Unit | Heating season | No heat. season | Whole year |
|---|---|---|---|---|
| NG demand | GWh | 31.65 | 4.37 | 36.02 |
| NG imported | GWh | 30.38 | 2.37 | 32.75 |
| | (%) | (96%) | (54%) | (91%) |
| SNG | GWh | 1.27 | 2.00 | 3.27 |
| | (%) | (4%) | (46%) | (9%) |

The demand for electricity increases in summer by about 10%, compared to the heating season, mainly due to the building cooling demand, while the production of RES increases by about 50%, thus creating much more RES over-generation (see Table 6). The RES peaks are higher than in winter, thus the RPF on the transformers increases (see the yellow areas in Fig. 10. About 80% of the energy surplus is absorbed by the P2G plants throughout the whole season. In particular, about 1.35 GWh of RPF is generated during the hot season, that is, 70% of the whole year's RPF. Of this, about 75% is generated on transformer #3.

The gas flows inside the network are much lower in the hot season than during the cold season (see Table 7 and Fig. 11a[1]). During the hot season, it could happen that the SNG production exceeds the gas demand of the network: when this happens, the SNG could be stored within the gas network volume. The gas network pressure increases, due to the linepack effect (see Fig. 11b). The SNG stored in the gas pipeline can be used later on, when the SNG production alone is not able to cover the network gas demand (see the white areas in Fig. 11a). The gas network can host the SNG production until its pressure is lower than the allowed maximum pressure (5 $bar_g$). In the case reported in Fig. 11, the network pressure reaches its upper constraint at 14:30, and the control system then limits the production of the methanation unit of P2G#1 and P2G#2 (see Fig. 12a). The control system choses to maintain the production of P2G#3 as, at that moment, the hydrogen buffer pressure of that plant is the highest of the three (see Fig. 12b). Even though the methanation unit of a plant is in standby, this does not affect the operation of the electrolyzer. For example, the methanation unit of P2G#2 remains in standby from 14:30 until 17:15 (see Fig. 12a); nevertheless, the electrolyzer continues to operate (see Fig. 10). The produced hydrogen is accumulated inside the buffer, whose pressure therefore increases (see Fig. 12b). Hydrogen can continue to be stored until the limit pressure of the buffer is reached. For example, at 15:45, the pressure of the P2G#1 buffer reaches its limit value and the electrolyzer must therefore limit its load (see Fig. 10).

[1] Note that for the sake of clarity, the scale used in Fig. 9a is different from that used in Fig. 11Fig. 11. Balance of the gas network (a) and gas network pressure (b) for the summer.
a.

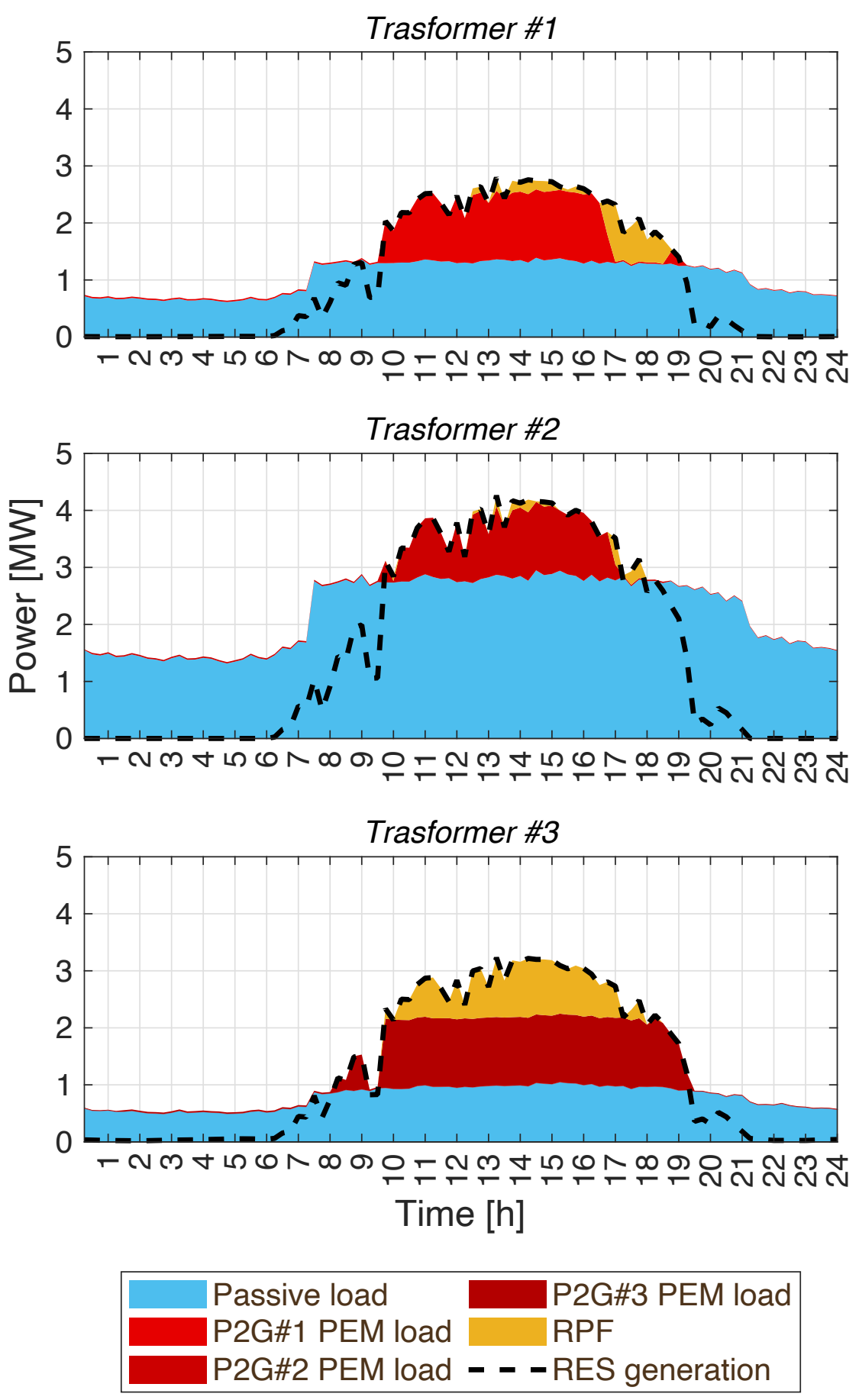

Fig. 10. Balance of the electricity network transformers for the summer.

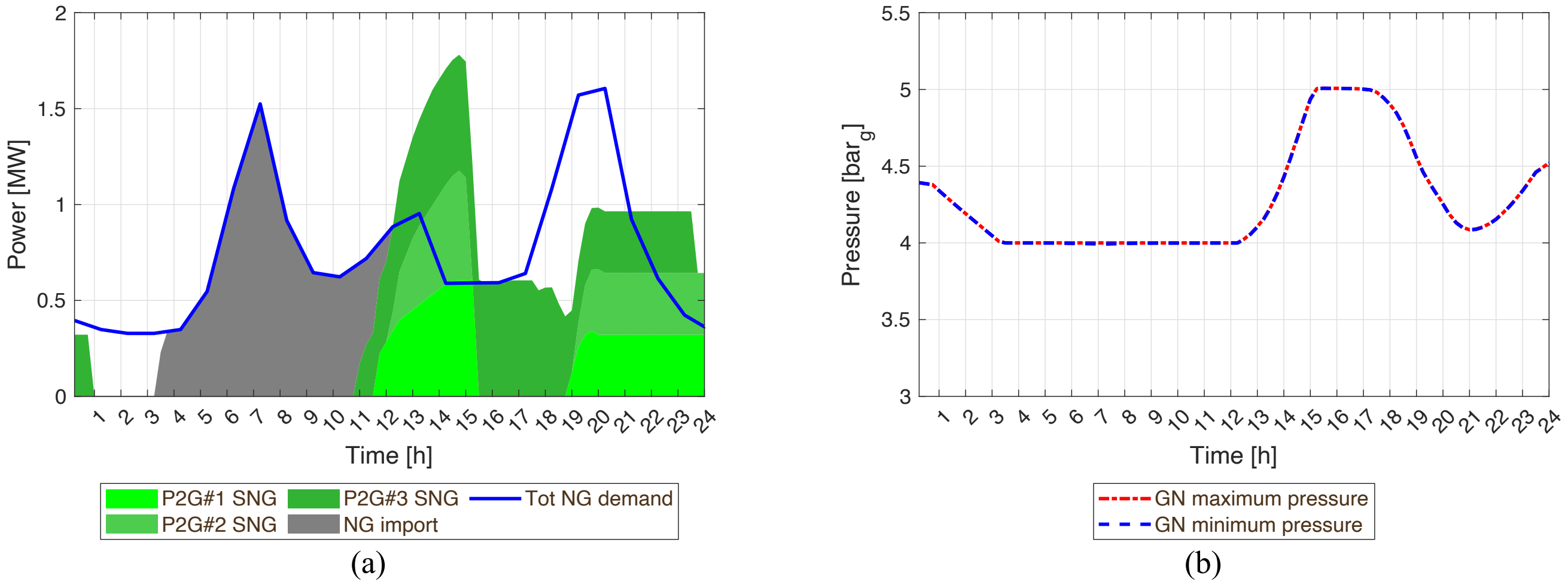

Fig. 11. Balance of the gas network (a) and gas network pressure (b) for the summer.

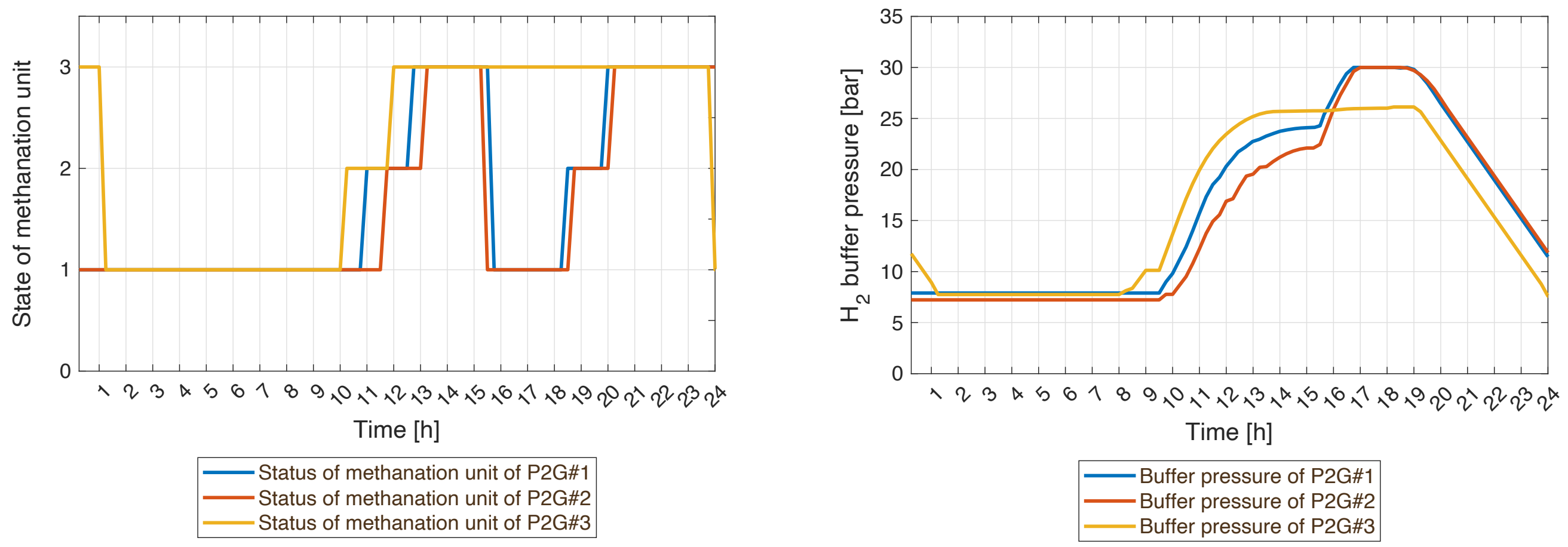

(a) (b)

Fig. 12. Status of the methanation units: 1 hot standby, 2 Reactor balancing phase, 3 up and running (a). Pressure of the hydrogen buffers (b).

3.2. Levelized cost of SNG

As previously mentioned, transformer #3 is the one that is affected by the highest RES overproduction. For this reason, P2G#3 is used more than the other two. This plant produces almost 3.5 GWh of SNG throughout the entire year, an amount which is approximately equal to the sum of the productions of the other two plants. Table 8 reports the production and consumption of the three plants: the higher the plant utilization is, the higher the production of SNG and of the indirect products, i.e., high temperature heat and oxygen. Considering the economic assumption to be equal, the higher the utilization of P2G is, the lower $LC_{SNG}$ (see Table 9). The $LC_{SNG}$ of P2G#3 is around 50-70% lower than the $LC_{SNG}$ of P2G#2 (the least used plant).
The scenario has been simulated under different economic assumptions: different cost scenarios, referring to years 2030 and 2050, and different RES surplus energy prices. This has allowed a sensitivity analysis, based on the costs and RES surplus prices, to be carried out.
The investment cost of the units that compose a P2G plant has been hypothesized to decrease by 2050, compared to 2030: a 25% cost reduction for the electrolyzer, a 10% reduction for the hydrogen buffer and a 33% reduction for the methanation reactor. This reduction leads to a reduction in $LC_{SNG}$, which decreases by 18-35%, depending on the different cases. Even when the cost of the RES surplus is hypothesized to be 0, $LC_{SNG}$ is not lower than the natural gas price. For this reason, green SNG will need more incentives to be able to compete with natural gas fossil fuel (in 2020, the average price of NG for non-household consumption in the European Union was 35 €/MWh [52]).
The results demonstrate that $LC_{SNG}$ depends to a great extent on both the economic assumption and on the plant operation and are in line with the results shown in the literature (see Fig. 13). The assumptions made in study [28] are similar to those made in this article: a 2050 scenario and a surplus electricity price of 8.71 €/MW. The cost of SNG for a plant size of around 1 MWe results be in the 200 to 350 € / MWh range. The analysis in [42] considers the cost of SNG for a 10 MWe plant, assuming a cost of electricity of 0 to 25 €/MWh: the SNG cost varies from 42 €/MWh to 313 €/MWh, as a function of the different configurations, when considering a 2030 scenario, and from 19 €/MWh to 170 €/MWh when considering a 2050 cost scenario. In [51], for a 10 MWe P2G plant with investment costs that refer to 2015 and an electricity cost of 40 €/MWh, $LC_{SNG}$ was around 170 €/MWh. Considering a 2030 scenario with an electricity cost equal to 60 €/MWh, $LC_{SNG}$ is 180 €/MWh, and $LC_{SNG}$ will be around 95 €/MWh in 2050, when considering an electricity price of 15 €/MWh. In [53], it was concluded that $LC_{SNG}$ will be in the 92 to 113 €/MWh range in 2050 when considering 25 €/MWh as the cost of electricity. In [54], it is reported that the SNG cost of a 10 MWe P2G plant would have been 55 €/MWh in 2020, considering a zero cost for electricity, and from 107 to 143 for an electricity price of 35 €/MWh. Considering the same electricity price, but for a 2030 scenario, the SNG cost will drop to 89-121 €/MWh and to 81-103 €/MWh in 2050.

Table 8. P2G results.

| | Unit | Heating season | | | | No heating season | | | | Whole year | | | |
|---|---|---|---|---|---|---|---|---|---|---|---|---|---|
| | | P2G#1 | P2G#2 | P2G#3 | Tot | P2G#1 | P2G#2 | P2G#3 | Tot | P2G#1 | P2G#2 | P2G#3 | Tot |
| El. cons | GWh | 0.72 | 0.50 | 1.69 | 2.90 | 1.32 | 0.98 | 2.14 | 4.44 | 2.03 | 1.48 | 3.83 | 7.34 |
| SNG | GWh | 0.29 | 0.19 | 0.79 | 1.27 | 0.58 | 0.40 | 1.02 | 2.01 | 0.87 | 0.59 | 1.81 | 3.27 |
| $CO_2$ | t | 59.8 | 38.5 | 159.4 | 257.6 | 119.4 | 84.7 | 204.7 | 408.8 | 179.2 | 123.2 | 364.1 | 666.5 |
| Heat | GWh | 0.09 | 0.06 | 0.23 | 0.37 | 0.17 | 0.12 | 0.30 | 0.59 | 0.26 | 0.18 | 0.53 | 0.96 |
| $O_2$ | t | 89.4 | 57.6 | 235.7 | 382.7 | 178.5 | 127.6 | 303.0 | 609.1 | 267.9 | 185.2 | 538.7 | 991.8 |

Table 9. Levelized cost of SNG.

| Electricity surplus price [€/MWh] | Levelized Cost of SNG ($LC_{SNG}$) [€/$MWh_{SNG}$] | | | | | |
|---|---|---|---|---|---|---|
| | 2030 | | | 2050 | | |
| | P2G#1 | P2G#2 | P2G#3 | P2G#1 | P2G#2 | P2G#3 |
| 0 | 166 | 255 | 70 | 97 | 153 | 37 |
| 5 | 177 | 266 | 81 | 108 | 164 | 47 |
| 15 | 199 | 290 | 101 | 130 | 188 | 68 |
| 30 | 232 | 325 | 132 | 163 | 223 | 99 |

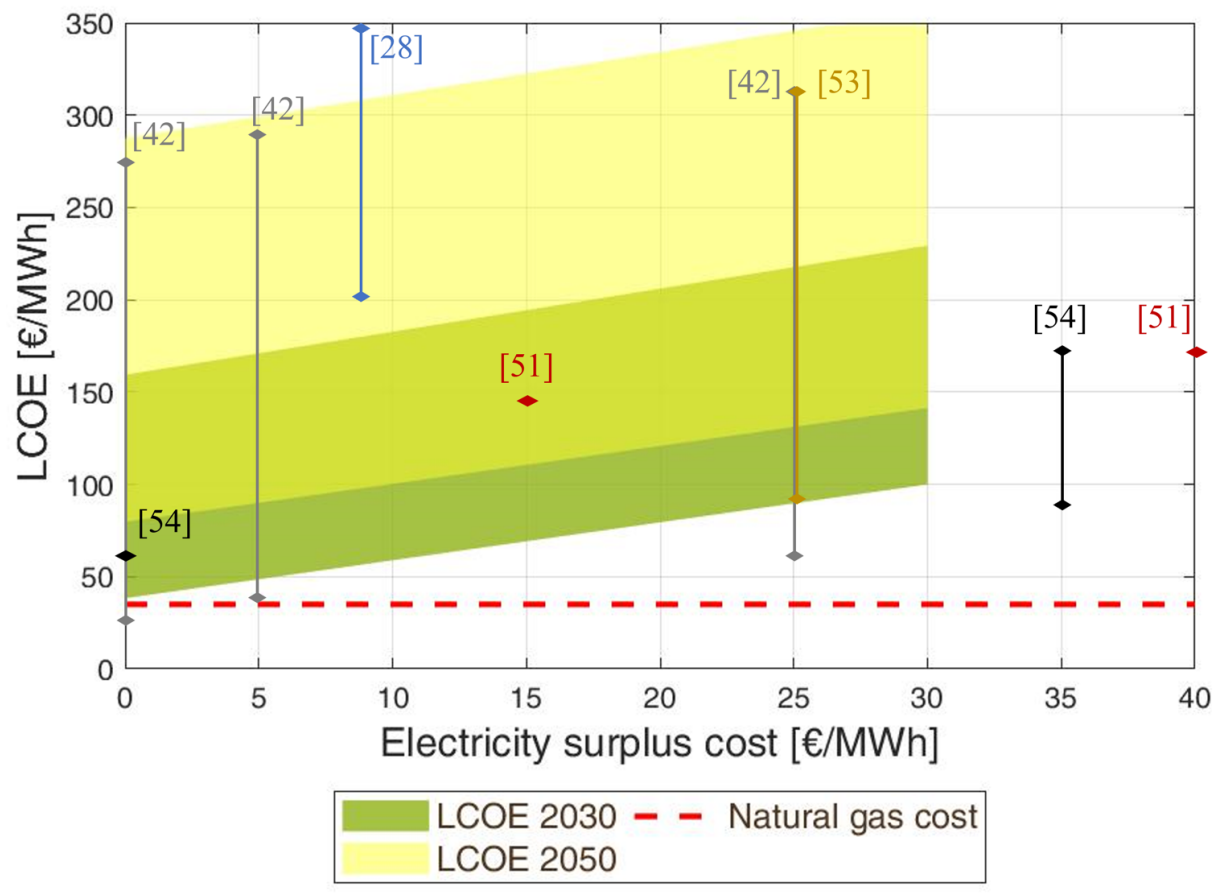

Fig. 13. Levelized cost of SNG.

### 3.3. Economic impact of the utilization of the heat and oxygen

The generation of oxygen and heat in a P2G plant is due to electrolysis and methanation thermochemical processes and they therefore represent by-products of the plant. As shown in Table 5, a remuneration of 30 €/MWh has been assumed for the produced high temperature heat and a remuneration of 70 €/t for the produced oxygen. The recovered heat can be sold for industrial use or even injected into the district heating network, if one exists [11], while oxygen could be sold for oxy-combustion in power plants or for medical care [55]. Vandewalle et al. [50] calculated that the valorization of produced oxygen could lead to a reduction in the SNG cost of as much as 20 €/MWh. This result is in line with what has been obtained in this study. In particular, the exploitation of produced oxygen leads to a reduction of 21 €/MWh in $LC_{SNG}$ for P2G#1 and P2G#3 and of 22 €/MWh for P2G#3. When the revenues from the heat recovery are also considered, a further cost reduction of SNG of around 9 €/MWh is obtained for all the plants. Considering the overall cost of the SNG, this cost reduction is not negligible: in the scenario analyzed in this paper, the heat and oxygen remuneration allowed $LC_{SNG}$ to be reduced by a minimum of 6% (in the P2G#2 case for the year 2030, the RES surplus price is 30 €/MWh) and a maximum of 36% (in the P2G#3 case for the year 2050, the RES surplus price is 0 €/MWh), with an average decrease of 15% when considering all the cases (see Fig. 14). Exploiting all the possible economic gains of the plant is of fundamental importance since, as has emerged in this analysis, the utilization of P2G technologies may lead to economic profitability problems.

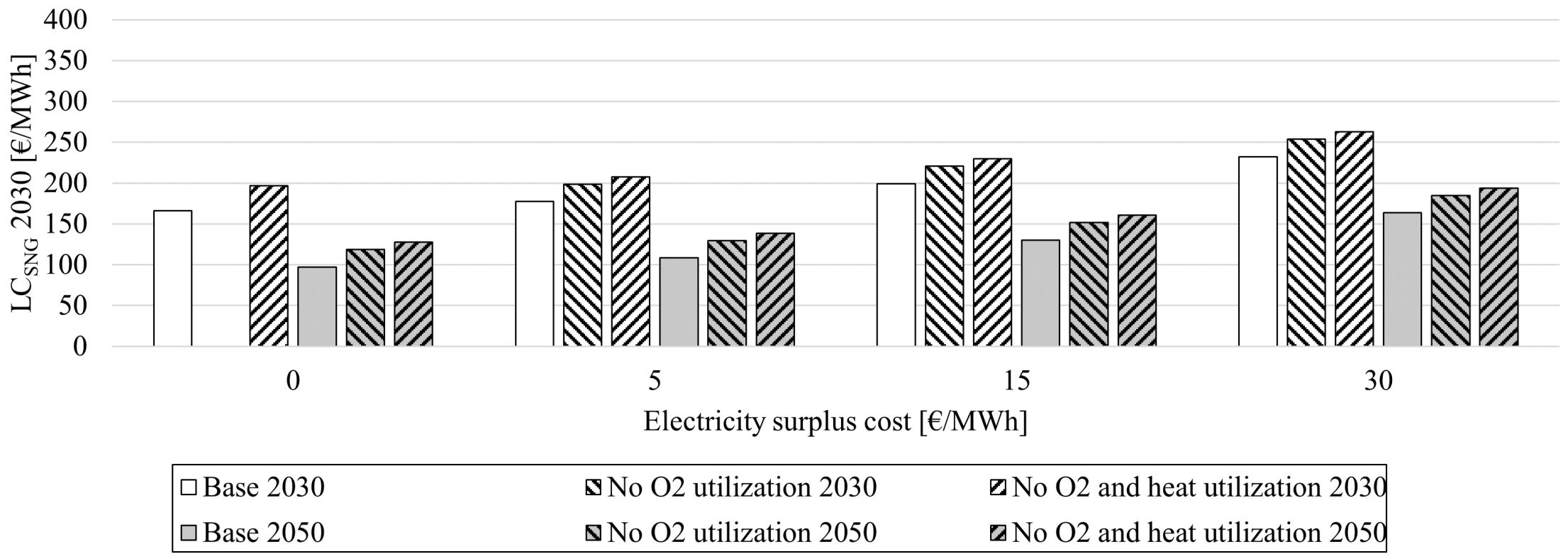

(a)

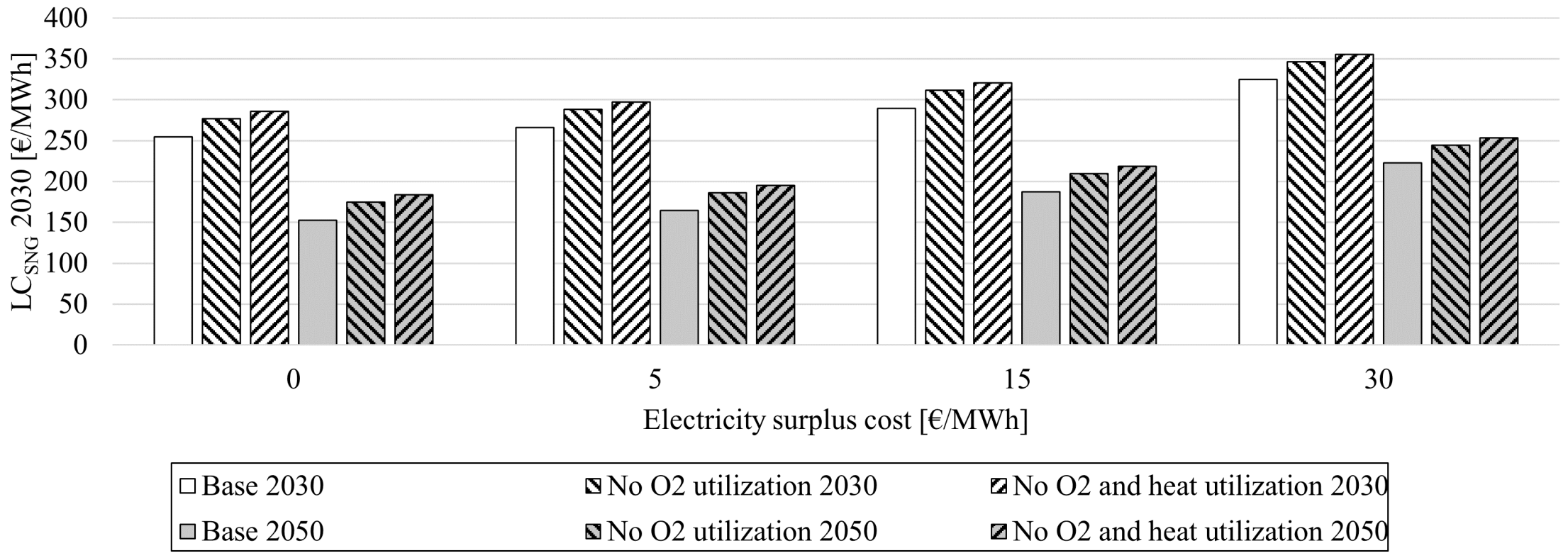

(b)

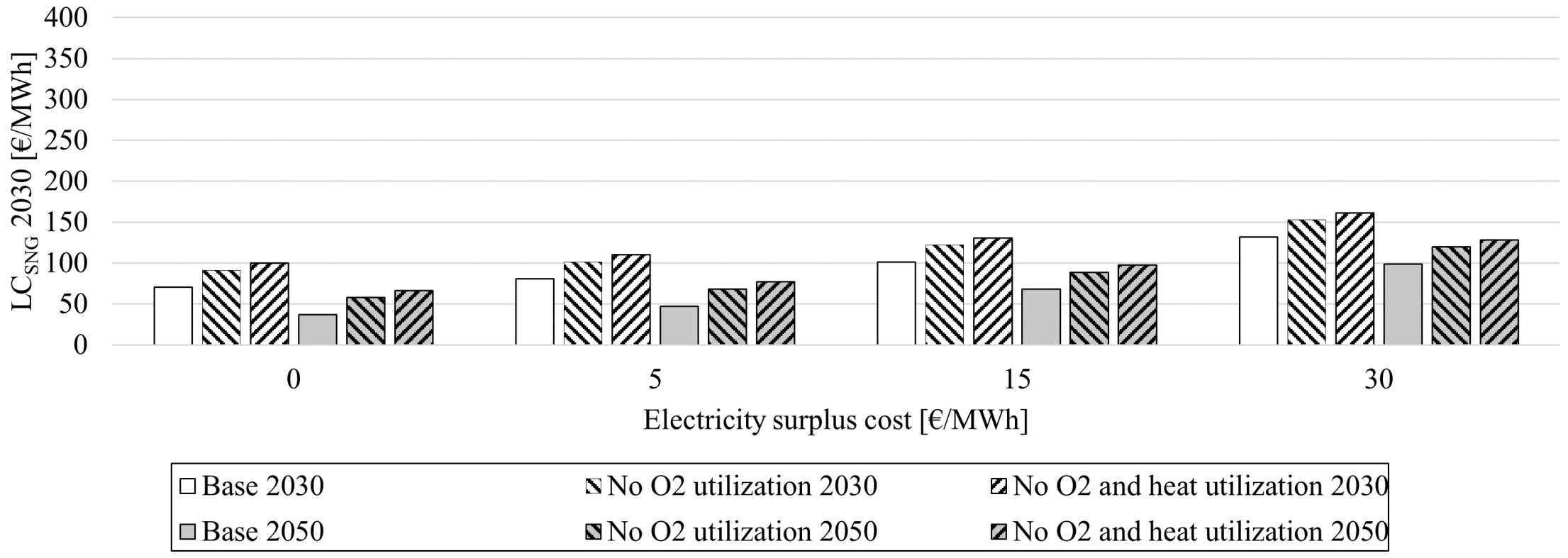

(c)

Fig. 14. Impact of oxygen and heat utilization for P2G#1 (a), P2G#2 (b) and P2G#3 (c).

## 4. Conclusions

A high penetration of non-dispatchable distributed generation may introduce issues and lead to challenges concerning the distribution systems. This paper has analyzed how P2G plants can be used to optimize the utilization of the overproduction of RES. The conclusion of this study can be summarized as follows:

- Thanks to the high flexibility of electrolyzer units, P2G plants can be effectively used to absorb the over-generation of renewable energy and thus alleviate the reverse power flow at the HV/MV transformers.
- The natural gas demand during the heating season is sufficiently high to allow SNG injection into the gas distribution network, without any kind of restriction. However, the P2G production in this season covers less than 5% of the gas demand. In the analyzed scenario, where the P2G plants are only used to absorb the overproduction of RES, the SNG production is not enough to decarbonize the gas sector.
- A low gas demand season may be critical for the utilization of P2G plants. In these periods, the SNG injection needs to be regulated to avoid over pressures in the gas network. This mainly happens in summer when there is no demand for gas for heating. In this season, the RES production is higher, and thus the criticality of the scenario increases. In these conditions, the P2G systems should be coordinated with each other to limit simultaneous gas injections. In such a context, different energy conversion systems, which share the same destination sink (in this case the gas network), need to cooperate with each other to optimize the utilization of the sink and maximize the benefits of the entre multi energy system.
- If the SNG production is not consumed directly (as happens during low gas demand periods), the gas distribution network can be used for gas storage purposes. When the SNG production is higher that the gas demand, the overproduction of SNG can be hosted inside the network volume until the pressure constraint is respected (linepack effect). It is well known that the linepack effect can also be used even for seasonal storage purposes in gas transmission networks. The lower pressure range of the distribution network (compared to the transmission one) does not allow seasonal storage in a medium pressure network but only intraday storage. Nevertheless, the distribution network linepack effect leads to greater flexibility in the use of the methanation units: SNG is accumulated within the gas network in the most critical periods of low gas demand and is then consumed in the following hours, thus allowing a more constant operation of the methanation units to be achieved.
- Even though the low gas demand may be a constraint for methanation units in some cases, it only has a limited impact on the electricity side. In fact, the hydrogen buffer of P2G plants allows utilization of the methanation unit and the electrolyzer to be decoupled. The RES overproduction is absorbed by the electrolyzer, the produced hydrogen can be accumulated in the buffer and then, at the opportune time, it is converted into SNG by the methanation unit and injected into the gas network.

- $LC_{SNG}$ varies from 49 to 319 €/MWh, according to the different assumptions and configurations.
- The cost of the units of the P2G plants for the year 2030 have been forecasted to decrease (a 25% cost reduction for the electrolyzer, a 10% reduction for the hydrogen buffer and a 33% reduction for the methanation reactor). This cost reduction is reflected on $LC_{SNG}$, which on average decreases by 30%, depending on the different configurations and assumptions.
- The cost of the RES energy surplus has been hypothesized to vary from 0 to 30 €/MWh$_e$ and this also impacts $LC_{SNG}$: the higher the cost of electricity is, the higher $LC_{SNG}$. Even when considering the maximum incentive (i.e., the cost of electricity equal to zero), $LC_{SNG}$ can be very high, especially if the plant is used to absorb small amounts of RES overproduction. Additional incentives are needed to make green SNG competitive with natural gas.
- The same scenario has been simulated considering the possible revenues from $O_2$ and the heat produced by the P2G plants. It has emerged that the exploitation of these indirect productions allows LCSNG to be reduced by around 30 €/MWh, which on average corresponds to 15% of the total LCSNG: it is thus important to take into consideration the possible incomes of these byproducts as the resulting gains may not be negligible.
- Even if all the plants are connected to the same electric distribution network, the location of the P2G plant has an important impact on $LC_{SNG}$. In order to increase the production of SNG, and therefore better amortize the investment costs, P2G needs to be installed in the part of the network where it is best able to absorb the overproduction of renewables.

As shown in the paper, the study of multi-energy systems requires a number of different competencies from different backgrounds. The use of conversion devices using different commodities (such as electricity and gas) calls for the modeling of different network infrastructures, because of the network constraints that may affect the operation of the conversion devices. The next works of the authors will focus on a comparison between different conversion systems (for example, Power-to-Gas and Power-to-Heat), to understand their contributions to the ongoing energy transition.

## Appendix A. Gas network model

The gas network model is based on the Renouard equation for the medium pressure [56] that defines the relation between the pressure difference between two nodes ($m$ and $n$) and the flow of natural gas in the pipe that connects these nodes ($m-n$):

$$P_m - P_n = \left[P_m - \sqrt{{P_m}^2 - 25.24 \cdot L_{m-n} \cdot \left(\frac{\dot{m}_{m-n} \cdot 3600}{\rho_{NG}}\right)^{1.82} \cdot {D_{m-n}}^{-4.82}}\right] \quad \text{(A1)}$$

where:

- $P_m$ and $P_n$ are the pressures of nodes $m$ and $n$, respectively [bar];
- $L_{m-n}$ is the length of pipe $m-n$ [m];
- $\dot{m}_{m-n}$ is the natural gas flow inside pipe $m-n$ [kg/s];
- $\rho_{NG}$ is the natural gas density under standard conditions [kg/m$^3$];
- $D_{m-n}$ is the dimeter of pipe $m-n$ [mm].

Then, the gas flow in a generic pipe $(m-n)$ $\dot{m}_{m-n}$ can be expressed as:

$$\dot{m}_{m-n} = \left|\frac{{P_m}^2 - {P_n}^2}{25.24 \cdot L_{m-n} \cdot {D_{m-n}}^{-4.82}}\right|^{\frac{1}{1.82}} \cdot \frac{\rho_{NG}}{3600} \cdot sgn(P_m - P_n) \quad \text{(A2)}$$

Under the hypothesis that the natural gas follows the ideal gas equation, the derivative of the pressure over time can be expressed as:

$$\frac{dP_m}{dt} = 10^5 \cdot \frac{R_{NG} \cdot T}{V_m} \cdot \dot{m}_m \quad \text{(A3)}$$

where:

- $P_m$ is the natural gas pressure [bar];
- $V_m$ is the volume of a node [m$^3$]
- $m_m$ is the mass [kg];
- $R_{NG}$ is the specific gas constant of natural gas [J/kg/K];
- $T$ is the natural gas temperature (considered to be constant) [K].

The node volume is assumed to be equal to half the sum of the volumes of all the $N$ pipes connected to the node:

$$V_m = \frac{\sum_{n=1}^{N} \left(\frac{D_{m-n}}{1000 \cdot 2}\right)^2 \cdot \pi \cdot L_{m-n}}{2} \quad \text{(A4)}$$

The continuity equation at node $m$ $is$:

$$\dot{m}_m = \dot{m}_{inj,m} - \dot{m}_{wit,m} + \sum_{n=1}^{N} \dot{m}_{m-n} \quad \text{(A5)}$$

where $\dot{m}_{inj,m}$ and $\dot{m}_{wit,m}$ are the gas injection and gas withdrawal at node $m$, respectively [kg/s];
Combining equations (A2), (A3), (A4) and (A5), it is possible to define the pressure variation of each node, according to the characteristics of the node, the pressures of the adjacent nodes and the gas injections and gas withdrawals in that node:

$$\frac{dP_m}{dt} = \frac{2 \cdot R_{NG} \cdot T}{\sum_{n=1}^{N} \left[\left(\frac{D_{m-n}}{1000 \cdot 2}\right)^2 \cdot \pi \cdot L_{m-n}\right]} \cdot \left\{ \dot{m}_{inj,m} - \dot{m}_{wit,m} + \sum_{n=1}^{N} \left[ \left| \frac{{P_m}^2 - {P_n}^2}{25.24 \cdot L_{m-n} \cdot {D_{m-n}}^{-4.82}} \right|^{\frac{1}{1.82}} \cdot \frac{\rho_{NG}}{3600} \cdot sgn(P_m - P_n) \right] \right\} \quad \text{(A6)}$$

The first node of the network is the city-gate, i.e., the connection point of the medium-pressure network to the high-pressure network. The pressure at this node is assumed to be constant:

$$P_1 = P_{citygate} \quad \text{(A7)}$$

If the pressure of node 2 (the downstream node of the city-gate) is lower than the city-gate pressure, the NG extracted from the city-gate is calculated using the Renouard correlation. Instead, if the pressure of node 2 is higher than the city-gate pressure, then the gas flow through the city-gate is 0: it is no possible to inject gas from a medium-pressure network to a high-pressure network.

$$\dot{m}_{citygate} = \dot{m}_{1-2} = \begin{cases} \left(\frac{{P_1}^2 - {P_2}^2}{25.24 \cdot L_{1-2} \cdot {D_{1-2}}^{-4.82}}\right)^{\frac{1}{1.82}} \cdot \frac{\rho_{NG}}{3600}, & if\ P_1 - P_2 > 0 \\ 0, & else \end{cases} \quad \text{(A8)}$$

At each time step, the GN module calculates the maximum amount of SNG that can be stored and then used for the control of the methanation units. The total amount of SNG that the network can accept is equal to the of gas withdrawals plus the amount of gas that can be accumulated as a result of the linepack effect: i.e., the quantity of gas which, if injected, would bring the network pressure to the maximum allowed pressure. The total amount of SNG that can be injected at generic timestep $t$ is defined as:

$$SNG_{max_t} = \left(\sum_{m=1}^{M} m_{wit,m_t}\right) + \frac{(P_{max} - \bar{P}_t) \cdot 10^5 \cdot V_{tot}}{R_{NG} \cdot T} \quad \text{(A9)}$$

where:

- $m_{wit,m_t}$ is the amount of NG withdrawn at node $m$ at time step $t$;
- M is the number of nodes;
- $P_{max}$ is the maximum allowed pressure in the network;
- $\bar{P}_t$ is the mean pressure in the network at time step $t$;
- $V_{tot}$ is the total volume of the network.

This simplified methodology has proved to be sufficiently accurate for the purposes of this study. The utilization of linepack could be further optimized, but this was beyond the scope of this paper.
In order to validate the model, the results have been compared with those obtained with the steady state and multi-component thermal-fluid-dynamic model presented by Cavana and Leone [35]. Their model is non-isothermal and considers NG as a gaseous hydrocarbon mixture. The comparison was made on a 78 node 4th species network, according to the Italian DM 24/ 11/1984 classification [36]. The relative error on the node pressure resulted to always be lower than 2% at each time step (see Fig. A.1).

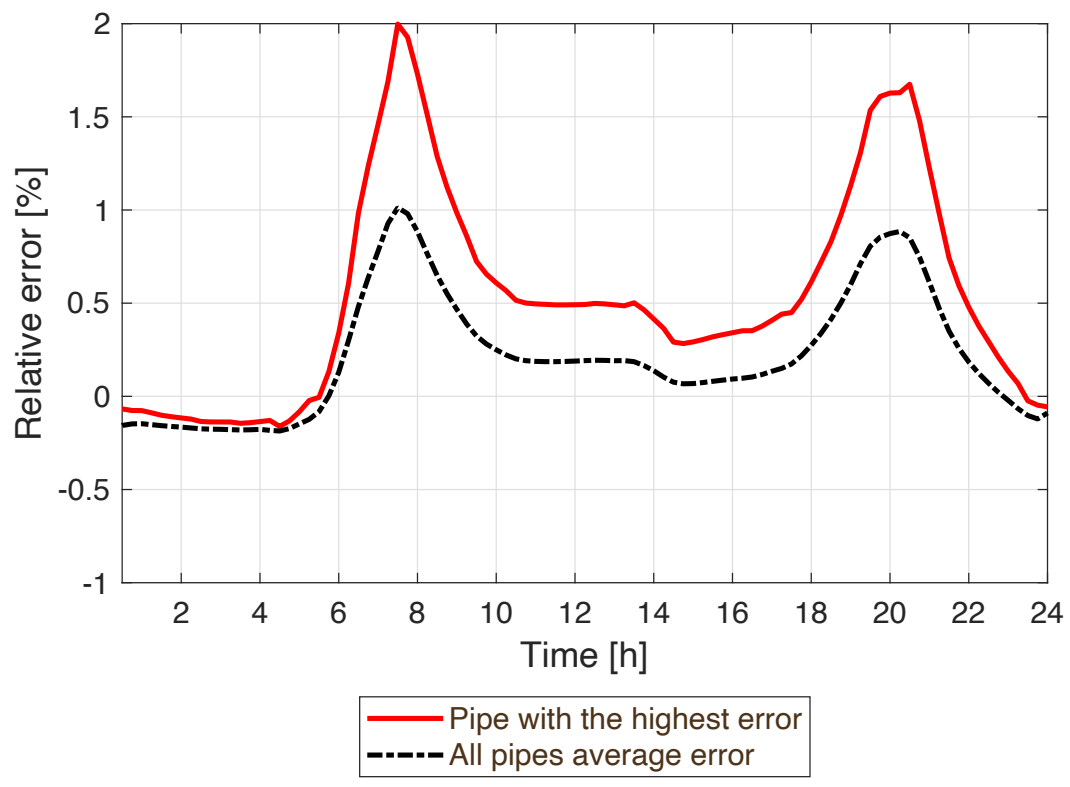

Fig. A.1. Relative error of the pressure in the network pipes: a pipe with the highest deviations and mean error (a day with the highest deviations).